\documentstyle[12pt,epsf,epsfig]{article}

\def\be{\begin{equation}}
\def\ee{\end{equation}}
\def\bea{\begin{eqnarray}}
\def\eea{\end{eqnarray}}

\begin{document}
\setlength{\baselineskip}{20pt}
\baselineskip 12pt

\newcommand{\sheptitle}
{Neutrino Oscillations: Status, Prospects and
Opportunities at a Neutrino Factory}
\newcommand{\shepauthor}
{S. F. King
\footnote{Based on invited talks at the IPPP Workshops
on Physics at a Future Neutrino Factory, 19 January 2001, RAL
and 21-23 March 2001, Durham. To appear in J.Phys.G}}

\newcommand{\shepaddress}
{Department of Physics and Astronomy,
University of Southampton, Southampton, SO17 1BJ, U.K.}

\newcommand{\shepabstract}
{We review the current status of neutrino oscillations
after 1258 days of Super-Kamiokande,
assess their future prospects over the next 10 years
as the next generation of experiments come on-line, and
discuss the longer-term opportunities presented by a Neutrino Factory.
We also give an introduction to the see-saw mechanism and 
its application to atmospheric and solar neutrinos.}

\begin{titlepage}
\begin{flushright}
hep-ph/0105261\\
\end{flushright}
\begin{center}
{\large{\bf \sheptitle}}
\bigskip \\ \shepauthor \\ \mbox{} \\ {\it \shepaddress} \\ \vspace{.5in}
{\bf Abstract} \bigskip \end{center} \setcounter{page}{0}
\shepabstract
\end{titlepage}

\section{Introduction}

The Super-Kamiokande experiment marks a watershed in neutrino
oscillation physics. Essentially Super-Kamiokande is a large
water Cherenkov detector consisting of 50,000 tons of pure water,
and 11,200 photomultiplier tubes, making it the largest detector
of its kind in the world. Neutrinos interact inside the detector
producing either electrons or muons which give rise
to the characteristic Cherenkov light cones which are observed
by the photo-tubes situated on the walls of the detector.
The results of the Super-Kamiokande experiment
\cite{SK} provides compelling evidence for neutrino oscillations
and hence neutrino mass, which if confirmed, would be the first
evidence for new physics beyond the Standard Model. 

The Super-Kamiokande
experiment has measured the number of electron and muon neutrinos
that arrive at the Earth's surface as a result of cosmic ray
interactions in the upper atomosphere, which are referred to as
``atmospheric neutrinos''. While the number and and angular
distribution of electron neutrinos is as expected, Super-Kamiokande
has shown that the number of muon neutrinos is significantly smaller
than expected and that the flux of muon neutrinos exhibits
a strong dependence on the zenith angle. These observations give
compelling evidence that muon neutrinos undergo flavour oscillations
and this in turn implies that at least one neutrino flavour
has a non-zero mass. The interpretation preferred by Super-Kamiokande
is that muon neutrinos are oscillating into tau neutrinos.
Several experiments that use neutrino beams from
particle accelerators will take data over the next few years. 
If these experiments confirm the Super-Kamiokande results then 
this would be the  first evidence that the Standard Model
is incomplete since it was written down. This has led to an explosion
activity in this area \cite{nu2000}, \cite{Reviews}.

Super-Kamiokande is also sensitive to the electron neutrinos
arriving from the Sun, the ``solar neutrinos'',
and has independently confirmed the reported deficit of such
solar neutrinos long reported by other experiments.
For example the Homestake Chlorine experiment which began data taking
in 1970 consists of 615 tons of tetrachloroethylene, and uses
radiochemical techniques to determine the Ar37 production rate.
More recently the SAGE and Gallex experiments contain large amounts
of Ga71 which is converted to Ge71 by low energy electron neutrinos
arising from the dominant pp reaction in the Sun.
The combined data from these and other experiments implies
an energy dependent suppression of solar neutrinos which
can be interpreted as due to flavour oscillations. Taken
together with the atmospheric data, this requires that a second
neutrino flavour has a non-zero mass. The minimal interpretation
is that the electron neutrinos oscillate into a linear combination
of muon and tau neutrinos.

There are a number of theories that can explain the
experimental observations. Many of them are based on the see-saw
mechanism \cite{seesaw,Models}. 
The see-saw mechanism assigns very heavy Majorana masses to the
right-handed partners of the left-handed neutrinos, in addition to the
usual Dirac masses which arise from the Yukawa couplings
of left-handed neutrinos to right-handed neutrinos in the
same way that charged lepton masses occur in the Standard Model.
However the heavy Majorana masses of the right-handed neutrinos
(which forbidden for the right-handed electron by electric
charge conservation)
leads to light physical Majorana neutrino masses via the
see-saw mechanism. With a suitable choice of masses of the heavy
Majorana masses, and Yukawa couplings, the masses and mixing angles
of the light neutrino states fall into ranges which are suitable
to accomodate the experimental observations. 

Other theories 
explain the tiny neutrino mass in terms of Supersymmetry
in which R-parity is violated, or in terms of extra dimensions. 
The R-parity explanations \cite{RPV} 
offer the prospect
of collider signatures which would confirm such explanations.
The large extra dimension approach \cite{LED}
predicts radical departures from
the oscillation picture which could be tested by the next generation
of neutrino experiments, and in addition this approach could
be tested by collider experiments.

Whatever the underlying mechanism, a measurement
of neutrino masses and mixing angles will  allow us to infer
information about the particular Grand Unified Theory
(GUT) or string theory that governs the
right-handed neutrinos. Furthermore, an accurate measurement
of these parameters will provide information in the poorly understood
lepton sector that will complement that already available in the 
quark sector. Such information will provide vital clues to the
problem of the origin of quark and lepton masses and may unlock
the puzzle of why there are three generations.

In section 2 we introduce neutrino masses and mixing angles.
Section 3 describes the current status of neutrino oscillations
while section 4 presents some simple patterns of neutrino masses which
can account for atmospheric and solar oscillations.
In section 5 we ask the question ``What will we know in 10 years
time?'', and in section 6 we discuss the longer term
opportunities for neutrino oscillation physics at a Neutrino Factory.
Section 7 introduces the see-saw mechanism and shows how
it may be applied to a hierarchical neutrino spectrum 
describing atmospheric and solar neutrino data using
single right-handed neutrino dominance, and section 8
concludes the paper.

\section{Neutrino Masses and Mixing Angles}
The minimal neutrino sector required to account for the
atmospheric and solar neutrino oscillation data consists of
three light physical neutrinos with left-handed flavour eigenstates,
$\nu_e$, $\nu_\mu$, and $\nu_\tau$, defined to be those states
that share the same electroweak doublet as the left-handed
charged lepton mass eigenstates.
Within the framework of three--neutrino oscillations,
the neutrino flavor eigenstates $\nu_e$, $\nu_\mu$, and $\nu_\tau$ are
related to the neutrino mass eigenstates $\nu_1$, $\nu_2$, and $\nu_3$
with mass $m_1$, $m_2$, and $m_3$, respectively, by a $3\times3$ 
unitary matrix introduced by Maki, Nakagawa and Sakata (MNS) \cite{MNS},
$U_{MNS}$,
\begin{equation}
\left(\begin{array}{c} \nu_e \\ \nu_\mu \\ \nu_\tau \end{array} \\ \right)=
\left(\begin{array}{ccc}
U_{e1} & U_{e2} & U_{e3} \\
U_{\mu1} & U_{\mu2} & U_{\mu3} \\
U_{\tau1} & U_{\tau2} & U_{\tau3} \\
\end{array}\right)
\left(\begin{array}{c} \nu_1 \\ \nu_2 \\ \nu_3 \end{array} \\ \right)
\; .
\end{equation}

Assuming the light neutrinos are Majorana,
$U_{MNS}$ can be parameterized in terms of three mixing angles
$\theta_{ij}$ and three complex phases $\delta_{ij}$.
A unitary matrix has six phases but three of them are removed 
by the phase symmetry of the charged lepton Dirac masses.
Since the neutrino masses are Majorana there is no additional
phase symmetry associated with them, unlike the case of quark
mixing where a further two phases may be removed.
The MNS matrix
may be expressed as a product of three complex Euler rotations,
\begin{equation}
U_{MNS}=U_{23}U_{13}U_{12}
\end{equation}
where
\begin{equation}
U_{23}=
\left(\begin{array}{ccc}
1 & 0 & 0 \\
0 & c_{23} & s_{23}e^{-i\delta_{23}} \\
0 & -s_{23}e^{i\delta_{23}} & c_{23} \\
\end{array}\right)
\end{equation}

\begin{equation}
U_{13}=
\left(\begin{array}{ccc}
c_{13} & 0 & s_{13}e^{-i\delta_{13}} \\
0 & 1 & 0 \\
-s_{13}e^{i\delta_{13}} & 0 & c_{13} \\
\end{array}\right)
\end{equation}

\begin{equation}
U_{12}=
\left(\begin{array}{ccc}
c_{12} & s_{12}e^{-i\delta_{12}} & 0 \\
-s_{12}e^{i\delta_{12}} & c_{12} & 0\\
0 & 0 & 1 \\
\end{array}\right)
\end{equation}

\begin{figure}
\centering
%\leavevmode
%\vspace{-2.cm}
\epsfxsize=6.in
\epsffile{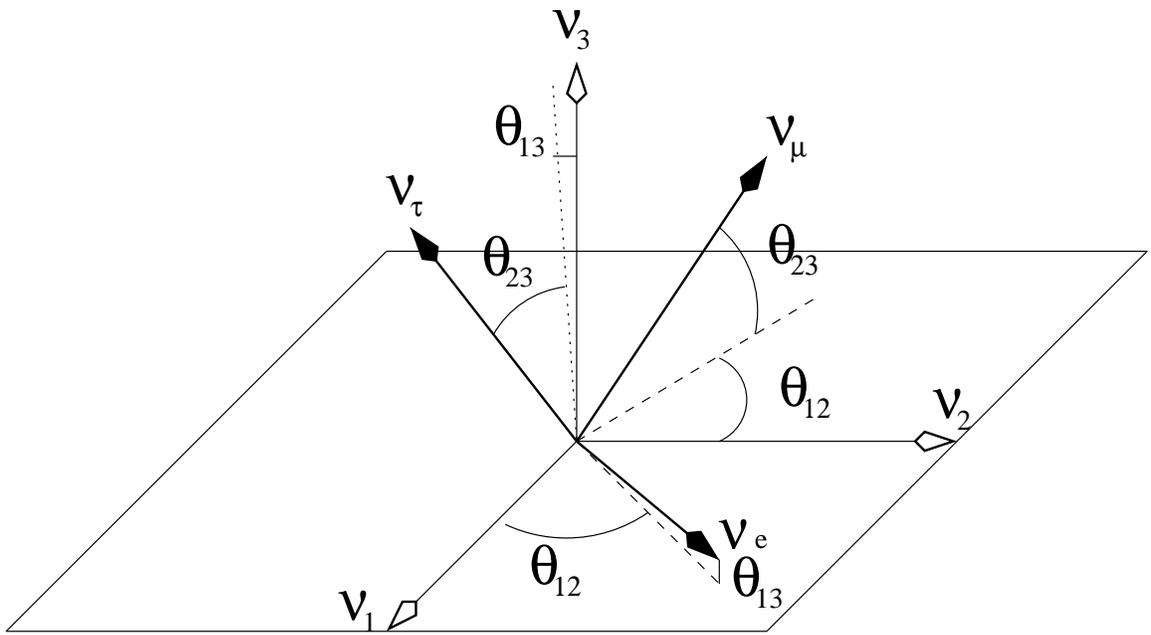}
%\vspace{2.cm}
\caption[]
{The relation between 
the neutrino flavor eigenstates $\nu_e$, $\nu_\mu$, and $\nu_\tau$ and
the neutrino mass eigenstates $\nu_1$, $\nu_2$, and $\nu_3$
in terms of the three mixing angles $\theta_{12}$,
$\theta_{13}$, $\theta_{23}$.}
\label{fig1}
\end{figure}

where $c_{ij} = \cos\theta_{ij}$ and $s_{ij} = \sin\theta_{ij}$.
Note that the allowed range of the angles is
$0\leq \theta_{ij} \leq \pi/2$. 
Since we have assumed that the neutrinos are Majorana, 
there are two extra phases, but only one combination
$\delta = \delta_{13}-\delta_{23}-\delta_{12}$
affects oscillations.
For the purposes of studying oscillation physics we may take
$\delta=\delta_{13}$, and $\delta_{23}=\delta_{12}=0$,
so that $U_{MNS}$ resembles the CKM matrix,
\begin{equation}
U_{MNS}=
\left(\begin{array}{ccc}
c_{12}c_{13} & s_{12}c_{13} & s_{13}e^{-i\delta} \\
-s_{12}c_{23}-c_{12}s_{23}s_{13}e^{i\delta}
& c_{12}c_{23}-s_{12}s_{23}s_{13}e^{i\delta}
& s_{23}c_{13} \\
s_{12}s_{23}-c_{12}c_{23}s_{13}e^{i\delta}
& -c_{12}s_{23}-s_{12}c_{23}s_{13}e^{i\delta}
& c_{23}c_{13} \\
\end{array}\right)
\end{equation}

It is convenient to define:
\begin{equation}
\Delta m^2_{ij} \equiv m^2_i - m^2_j \; .
\end{equation}
Oscillation probabilities depend upon the time--of--flight (and hence the
baseline $L$), the $\Delta m^2_{ij}$, and $U_{MNS}$ (and hence
$\theta_{12}, \theta_{23}, \theta_{13}$, and $\delta$).

Ignoring phases, the relation between 
the neutrino flavor eigenstates $\nu_e$, $\nu_\mu$, and $\nu_\tau$ and
the neutrino mass eigenstates $\nu_1$, $\nu_2$, and $\nu_3$
is just given as a product of three Euler rotations as
depicted in Figure \ref{fig1}.

\section{Status of Neutrino Oscillations}

Current atmospheric neutrino oscillation data are well described
by simple two-state mixing
\begin{equation}
\left(\begin{array}{c} \nu_\mu \\ \nu_\tau \end{array} \\ \right)=
\left(\begin{array}{cc}
 c_{23} & s_{23} \\
 -s_{23} & c_{23} \\
\end{array}\right)
\left(\begin{array}{c}
\nu_2 \\ \nu_3 \end{array} \\ \right)
\; ,
\end{equation}
and the two-state Probability oscillation formula
\be
P(\nu_{\mu }\rightarrow \nu_{\tau})=\sin^22\theta_{23}
\sin^2(1.27\Delta m_{32}^2{L}/{E})
\ee
where $\Delta m_{32}^2$ is in units of eV$^2$, the baseline $L$ is in km and
the beam energy $E$ is in GeV.
The atmospheric data is statistically dominated by the
Super-Kamiokande results and the latest 1117 day data sample leads
to \cite{nu2000}:
\begin{itemize}
\item $\sin^22\theta_{23}>0.88$
\item $1.5\times 10^{-3}$ eV$^2 < |\Delta m_{32}^2| 
<5\times 10^{-3}$ eV$^2$ (90\% CL)
\end{itemize}

CHOOZ data from $\bar{\nu}_{e}\rightarrow \bar{\nu}_{e}$
disappearance not being observed
\cite{chooz} provides a significant constraint on
$\theta_{13}$ over the Super-Kamiokande (SK) prefered
range of $\Delta m_{32}^2$:
\begin{itemize}
\item $\sin^22\theta_{13}<0.1-0.3$
\end{itemize}

\begin{figure}
\centering
%\leavevmode
%\vspace{-2.cm}
\epsfxsize=4.in
\epsffile{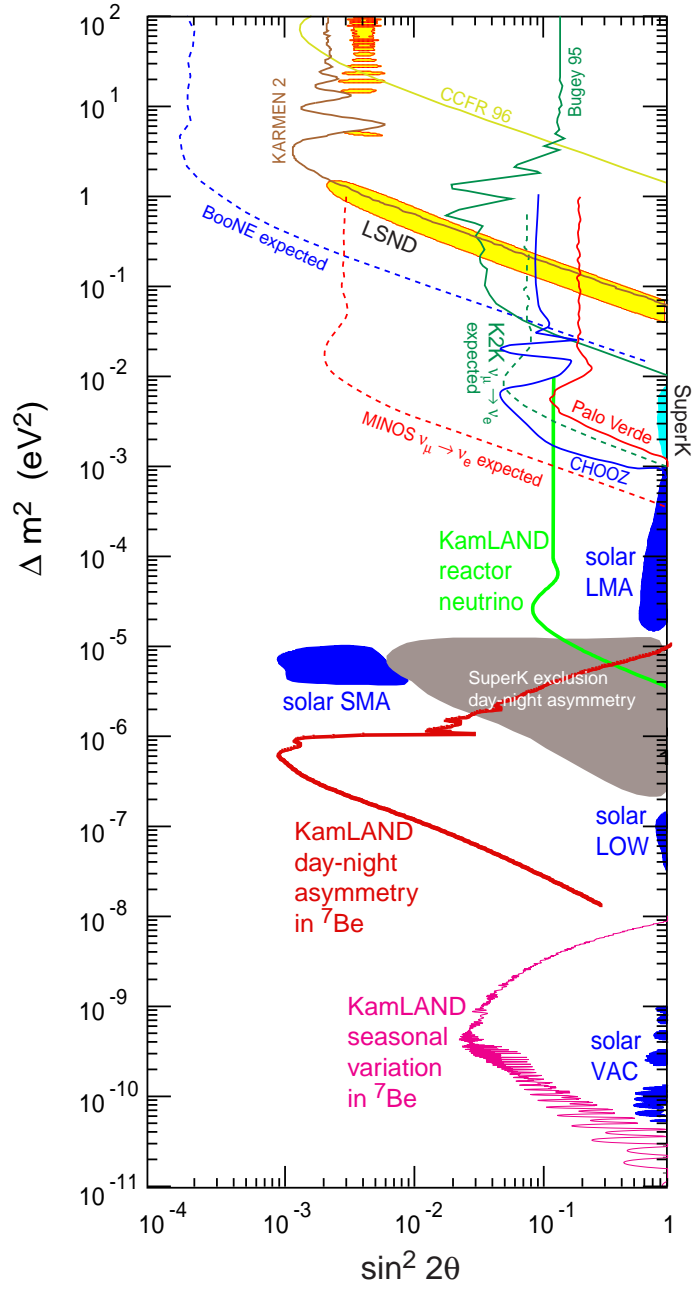}
%\vspace{2.cm}
\caption[]
{Summary of all oscillation data (taken from \cite{PDG}).}
\label{fig2}
\end{figure}

The solar neutrino problem can
be solved by four different combinations of oscillation
parameters, three of which are based on the MSW mechanism
\cite{MSW}, namely the large mixing angle (LMA), the
small mixing angle (SMA) and the long wavelength (LOW)
solutions, and a fourth solution represents 
vacuum (VAC) oscillations \cite{VAC}.
Various groups have performed fits to the
solar and atmosheric solutions \cite{atm99,MSW99,Barger:1999sm}.
Representative values of the parameters associated with
the four solutions based on the 
most recent data are given in Table 1 \cite{concha}.
\begin{table}[tbp]
\hfil
\begin{tabular}{ccc}
\hline \hline 
         &      $\sin^22\theta_{12}$  & $\Delta m_{21}^2\ \ (eV^2)$
         \\ \hline \hline 
LMA & 0.78 & $3.3\times 10^{-5}$ \\ \hline
SMA & 0.0027 & $5.1\times 10^{-6}$ \\ \hline
LOW & 0.93 & $9.6\times 10^{-8}$\\ \hline
VAC & 0.93 & $8 \times 10^{-10}$\\ \hline \hline 
\end{tabular}
\hfil
\caption{\footnotesize Typical best fit solar solutions 
(from \cite{concha}). }
\end{table}

In Figure \ref{fig2} we give a simplified summary of all 
oscillation data, based on two-flavour analyses,
and ignoring the ``dark side'' of the parameter space
corresponding to assuming that $\theta_{ij} \leq \pi/4$
\cite{darkside}. For large angle mixing which depends on matter
enhancement effects it becomes a relevant
question whether $\theta_{ij} < \pi/4$ or $\theta_{ij} > \pi/4$,
but the VAC solution is symmetrical about $\theta_{12} = \pi/4$.
This figure, taken from \cite{PDG} 
also includes a signal of neutrino
mass from the LSND experiment \cite{LSND}.
If confirmed this would require
the introduction of a fourth physical neutrino flavour, in order
to achieve three different mass squared splittings.
Since LEP data from the Z boson width only allows three
light neutrinos, the fourth light neutrino would have to be
a sterile neutrino which does not carry any quantum numbers,
unlike the three sequential neutrinos which carry weak
quantum numbers. We shall not consider the implications of the LSND result
any further here. 

\begin{figure}
\centering
%\leavevmode
%\vspace{-2.cm}
\epsfxsize=4.in
\epsffile{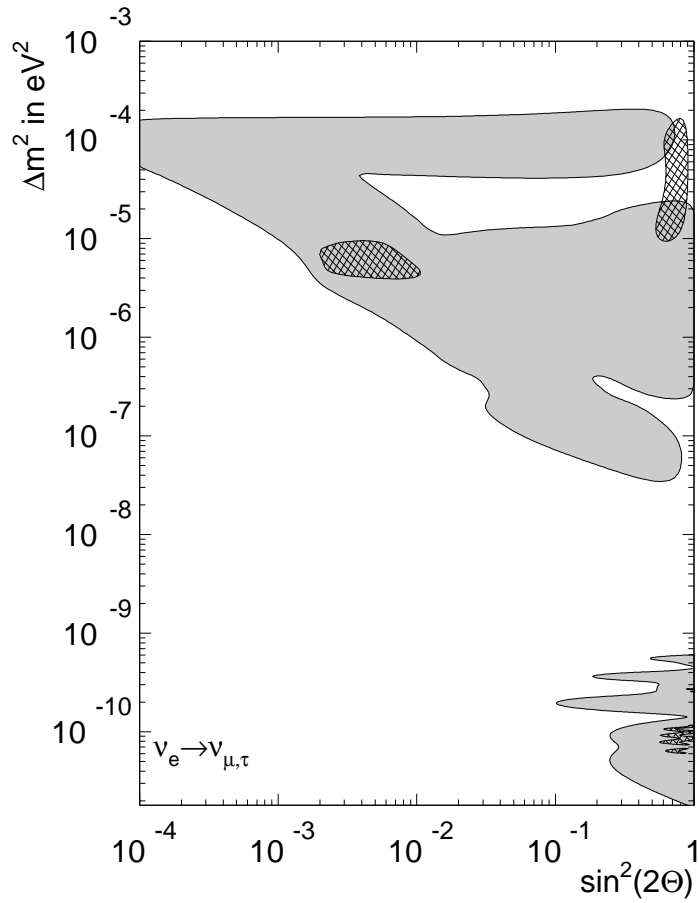}
%\vspace{2.cm}
\caption[]
{The Super-Kamiokande day-night exclusion area (shaded) 
based on 1258 days of data
compared to the flux allowed area (hatched) from all experiments
\cite{SKsolar}.}
\label{SKsolar}
\end{figure}

Recently Super-Kamiokande have reported constraints on neutrino
oscillations using 1258 days of their solar neutrino data \cite{SKsolar}.
They do not see any significant energy spectrum distortion, or
day/night effect, leading to strong constraints on the allowed
regions as shown in Figure \ref{SKsolar}. 
They claim that for active neutrino
oscillations, the SMA and VAC solutions are disfavoured at 97\% C.L.
from the (absence of the) day/night effect, 
and sterile neutrino oscillations are disfavoured at 95\% C.L.
They conclude that LMA active neutrino oscillations are preferred.
The results in Table 1 \cite{concha} are also based on 
an analysis performed on a similar data set, but using
a different global analysis technique, and concludes that
all four solar solutions remain viable although LMA is slightly
preferred. Part of the confusion is due to the fact that it is the
absence of any ``smoking guns'' for neutrino oscillations
which is leading to the constraints.
Interestingly if only the Super-Kamiokande data is considered,
and the other solar neutrino experiments are removed from the
analysis, then a large region of the LOW solution opens up
for both active and sterile neutrinos since there would
be no ``smoking gun'' sensitivity from either the day/night
effect or the spectral distortion \cite{SKsolar} in this region.

To summarise the current experimental situation,
it is probably premature to exclude any of the four solar
solutions in  Table 1. We can be reasonably confident that 
$\theta_{23}\approx \pi/4$, $\theta_{13}\leq 0.1$, while
$\theta_{12}$ might be either large or small.
Although nothing is known about the absolute scale
or ordering of the
mass squareds $m_i^2$, we do know that the ratio
\begin{equation}
R\equiv |\Delta m^2_{21}|/|\Delta m^2_{32}|
\equiv |\Delta m^2_{sol}|/|\Delta m^2_{atm}|
\label{R}
\end{equation}
shows a hierarchy which,
however, depending on the solar solution could be rather mild,
or relatively strong:
\begin{itemize}
\item $R={\cal O}(10^{-2})$ (LMA)
\item $R={\cal O}(10^{-7})$ (VAC)
\end{itemize}

\section{Patterns of Neutrino Masses}

\begin{figure}
\centering
%\leavevmode
%\vspace{-2.cm}
\epsfxsize=6.in
\epsffile{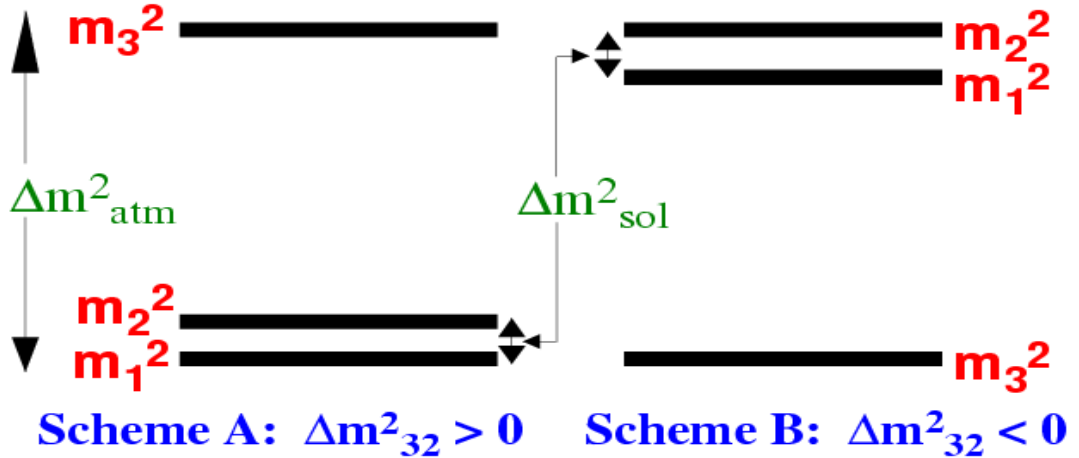}
%\vspace{2.cm}
\caption[]
{Alternative neutrino mass patterns that are consistent with neutrino
oscillation explanations of the atmospheric and solar neutrino
deficits
(taken from ref.\cite{Geer}.)}
\label{fig3}
\end{figure}

There are two simple patterns of neutrino masses which 
can account for the atmospheric and solar neutrino data,
as shown in Figure \ref{fig3}. These are based on either 
the hierarchy $|m_3|\gg |m_2| \gg |m_1|$ (scheme A) 
or the inverted hierarchy 
$|m_1|\approx |m_2| \gg |m_3|$ (scheme B).
In the case of a three neutrino hierarchical spectrum (scheme A),
there will be one
physical neutrino with a mass of about $|m_3|\approx 5\times 10^{-2}$ eV,
which would contribute as much ``hot'' dark matter
to the universe as all the visible
stars, and so would be of some relevance for cosmological structure formation.
A second
physical neutrino with a mass of about $|m_2|\approx 5\times 10^{-3}$
eV must also exist in order to account for the solar oscillation data.
In the case of an inverted hierarchy 
\cite{inverted} (scheme B) there are two 
neutrinos each with a mass of
about $|m_1|\approx |m_2|\approx 
5\times 10^{-2}$ eV but with a small mass splitting
of order $2\times 10^{-4}$ eV,
giving twice as much ``hot'' dark matter.
Therefore knowing the neutrino spectrum would be of some interest for
cosmological dark matter even in these minimal scenarios,
although the amount of dark matter in both cases is very small.
The inverted spectrum is technically natural,
in the sense of being stable under radiative corrections,
and predicts a large solar angle \cite{inverted}.

In addition a third pattern of neutrino masses has been proposed
(scheme C) based on all three neutrinos being approximately degenerate
with a mass of order an eV, but with very small mass splittings
suitable to describe the atmospheric and solar data.
In this approximately degenerate case there could be a substantial
component of ``hot'' dark matter, although in general such a pattern
of neutrino masses tends to be unstable under radiative
corrections unless protected by a symmetry \cite{Ellis}.

\section{What Will We Know in 10 years Time?}

The answer to this question is not completely clear,
which is one of the main reasons why neutrino physics
is so exciting at the current time.
In the near future much better solar neutrino measurements 
will be available as SNO, KamLAND, and  Borexino furnish us
with new data \cite{SNO,Kam,Bor}. For example,
Borexino is a new solar neutrino experiment at Gran Sasso
which will measure the flux of the low energy Be7 neutrinos in real time,
which is expected to be suppressed according to the MSW solutions.
Taken together these experiments will provide
``smoking gun'' evidence for neutrino oscillations, if it exists,
in all the currently allowed solar parameter ranges.

The Sudbury Neutrino Observatory (SNO) experiment is presently taking
data and the first results are expected very soon.
The SNO detector contains 1000 tonnes of heavy water D2O
and 9600 photo tubes. The use of heavy water, especially when 
enriched with salt, will enable the neutral current
(Z exchange) interactions of the neutrinos to be measured,
and hence the total active neutrino flux to be measured.
Thus SNO will see an excess of neutral current events
for flavour oscillations irrespective of the values taken by the
mixing parameters. A combination of spectral distortions,
day-night rate differences, and seasonal rate differences in the 
above experiments will then be used to distinguish betwen the
currently allowed parameter ranges. 

KamLAND (Kamioka Liquid scintillator Anti-Neutrino Detector)
will study antineutrinos produced by nearby nuclear reactors,
and will for example cover the LMA MSW region ``in the laboratory''.
Of particular interest
to the Neutrino Factory proposal is the ability of
KamLAND to confirm neutrino oscillations with a terrestrial
experiment if the parameters lie in the LMA range necessary for 
CP violation to be observed at a Neutrino Factory (as discussed
later). It is difficult to say how accurately $\theta_{12}$
and $\Delta m^2_{21}$ will be measured as this depends on the 
values of the other parameters, however an accuracy of $\sim 10-20\%$
seems possible. It is also possible that within 10 years we would have
a real-time measurement of the pp solar neutrino spectrum from
SIREN, LENS or HELLAZ \cite{SIREN,LENS,HELLAZ}. 
This would improve the accuracy of our parameter determination.

Over the next ten years the long baseline (LBL) experiments such as
K2K, MINOS and eventually the  CERN to Gran Sasso
experiments will report new results \cite{K2K,MINOS,CERN}.
The K2K experiment (a neutrino beam from from KEK to Super-Kamiokande) 
has already reported results from its first two years
of running \cite{K2K}.
These experiments are capable of confirming or rejecting the
hypothesis that neutrino oscillations are required to
explain Super-Kamiokande atmospheric neutrino results.
If the Super-Kamiokande results are confirmed then $\theta_{23}$
and $\Delta m^2_{32}$ may be measured with an accuracy of about $\sim
10\%$. MINOS will be sensitive to $\sin^22\theta_{13}$
values as low as 0.01 \cite{MINOS}.
In addition to these LBL experiments there is a proposal to
build a ``super'' neutrino beam line at the recently approved
Japanese Hadron Facility (JHF) \cite{JHF}. This beam would illuminate
the Super-Kamiokande detector, and enable a more accurate
determination of $\sin^22\theta_{13}$.
Discussions have also begun
\cite{Fermi} on building such a high-flux beamline at Fermilab.
In addition, it is proposed to use the proton driver for the
CERN Neutrino Factory, the SPL,  to provide a low energy neutrino
beam, pointing at the  Modane laboratory in the Alps. This
will form the first stage of the CERN Neutrino Factory project.
However, the precision of these experiments will be limited by the
uncertainty in the beam flavour content. For example, K2K
currently quote an error of 7\%, while MINOS hopes to reduce
this to 2\%. Such super-beam projects may be regarded as 
a natural stepping stone to a Neutrino Factory, to which we
now turn.

\section{Opportunities at a Neutrino Factory}

What is a Neutrino Factory? Basically it is a high intensity
muon storage ring with long straight sections along which the muons
decay to deliver a high intensity beam of neutrinos \cite{NF1,NF2,NF3}.
A typical high performance Neutrino Factory would involve
50 GeV muon beams and would deliver about $10^{20}$ muon decays per year.
The resulting neutrino beams are clearly of very high energy and
intensity and have precisely predictable neutrino flavour content,
making them superior to the conventional neutrino beams 
such as those proposed for example in \cite{JHF,Fermi}.

Despite the 
anticipated progress, even in 10 years time there are three quantities
which will remain obscure even in the most optimistic scenarios,
and these quantities would be measurable at a Neutrino Factory.
We shall refer to them as the three missing si(g)ns:
\begin{center}
\begin{enumerate}
\item The CHOOZ angle $\sin^22\theta_{13}$ which will still be
ill-determined even after LBL experiments, and is measureable
down to an astonishing 0.0001 at a Neutrino Factory.
\item The CP violating phase $\sin \delta$ which is impossible to
measure unless we have the LMA MSW solution.
\item The sign of the 23 mass splitting $(\Delta m_{32}^2)$,
which remains ambiguous as shown in Figure \ref{fig3},
is easy to measure at a Neutrino Factory.
\end{enumerate}
\end{center}

\begin{figure}
\centering
%\leavevmode
%\vspace{-2.cm}
\epsfxsize=4.0in
\epsffile{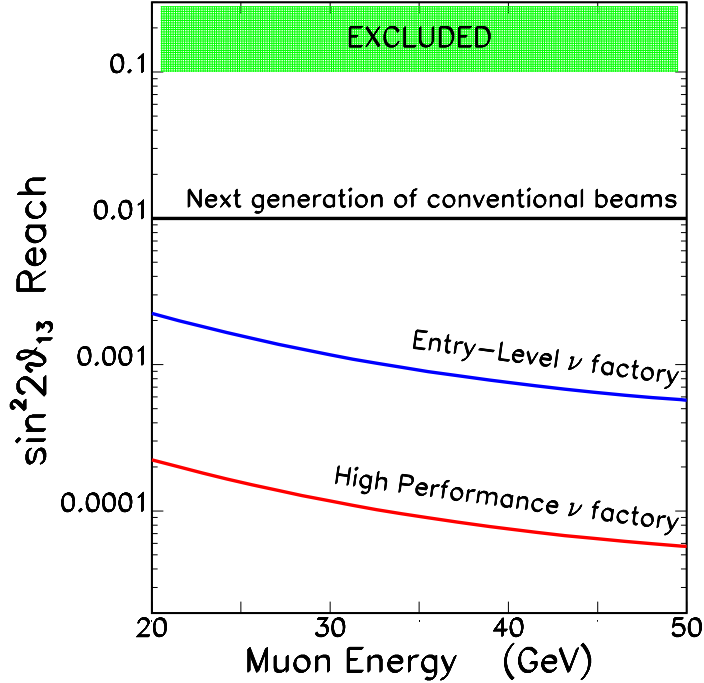}
%\vspace{2.cm}
\caption[]
{Reach in $\theta_{13}$ at a Neutrino Factory
(taken from ref.\cite{Geer}.)}
\label{fig4}
\end{figure}

It is worth emphasising that an accurate measurement of
$\sin^22\theta_{13}$ will be important
in discriminating GUT and string theories \cite{HK},
and that it if this angle is smaller than 0.01 it will simply be unmeasured
by the LBL experiments. At a Neutrino Factory it is relatively
straightforward to measure this angle using the Golden Signature
of ``wrong sign'' muons. For example suppose there are positive
muons circulating in the storage ring, then these decay as
$\mu^+ \rightarrow e^+\nu_e \bar{\nu}_\mu$ giving a mixed beam
of electron neutrinos and muon anti-neutrinos. The
muon anti-neutrinos will interact in the far detector to produce
positive muons. Any ``wrong sign'' negative muons which
may be observed can only arise from the neutrino oscillation
of electron neutrinos into muon neutrinos with probability given
by  a CP conserving part $P^{+}$ and a  a CP violating part
$P^{-}$. The exact formulae in vacuum are given by:
\be
P(\nu_{e}\rightarrow \nu_{\mu})= P^+(\nu_{e}\rightarrow \nu_{\mu})+
P^-(\nu_{e}\rightarrow \nu_{\mu})
\ee
where the CP conserving part is
\bea
P^+(\nu_{e}\rightarrow \nu_{\mu})  =
& - & 4Re(U_{e1}U_{\mu 1}^*U_{e2}^*U_{\mu 2})
\sin^2(1.27\Delta m_{21}^2L/E) \nonumber \\
& - & 4Re(U_{e1}U_{\mu 1}^*U_{e3}^*U_{\mu 3})
 \sin^2(1.27\Delta m_{31}^2L/E) \nonumber \\
& - & 4Re(U_{e2}U_{\mu 2}^*U_{e3}^*U_{\mu 3})
\sin^2(1.27\Delta m_{32}^2L/E)
\label{P+}
\eea
and the CP violating part is
\bea
& & P^-(\nu_{e}\rightarrow \nu_{\mu})  =
 - c_{13}\sin2\theta_{13}\sin2\theta_{12}\sin2\theta_{23}\sin \delta
\nonumber \\
& \times & \sin (1.27\Delta m_{21}^2L/E) \sin (1.27\Delta m_{31}^2L/E)
\sin (1.27\Delta m_{32}^2L/E)
\label{P-}
\eea
Note that $P^-$ requires all three families to contribute, and it
vanishes if any mixing angle or mass splitting is zero.
The angle $\theta_{13}$ may easily be extracted from
$U_{e3}$ in the dominant CP conserving term $P^+$, leading to
the expected limits in this angle at a Neutrino Factory
as shown in Figure 
\ref{fig4}.

In order to determine the CP violating phase $\sin \delta$
it is necessary to measure the CP violating term $P^-$.
In order to do this one must compare the result for
$P(\nu_{e}\rightarrow \nu_{\mu})$ to the result to the case where
the positive muons in the storage ring are replaced by negative muons
and the analagous experiment is performed to measure
$P(\bar{\nu}_{e}\rightarrow \bar{\nu}_{\mu})$.
The CP violating asymmetry due to the CP violating phase $\delta$
is given by
\be
A^{\delta}= \frac{P(\nu_{e}\rightarrow \nu_{\mu})-P(\bar{\nu}_{e}\rightarrow \bar{\nu}_{\mu})}
{P(\nu_{e}\rightarrow \nu_{\mu})+P(\bar{\nu}_{e}\rightarrow \bar{\nu}_{\mu})}
\label{ACP}
\ee
from which we obtain
\be
A^{\delta}= \frac{P^-(\nu_{e}\rightarrow \nu_{\mu})}
{P^+(\nu_{e}\rightarrow \nu_{\mu})}
\approx \frac{\sin 2\theta_{12}\sin \delta}{\sin \theta_{13}}
\sin (1.27\Delta m_{21}^2L/E)
\ee
It is clear that in order to measure the CP asymmetry we require large
$\theta_{12}$ and large $\Delta m_{21}^2$ and this corresponds to
the LMA MSW solution. In addition we require large $\sin \delta$.
Also it would seem that having small $\theta_{13}$ enhances the CP
asymmetry, however it should be remembered that the CP asymmetric
rate $P^-$ in Eq.\ref{P-} is proportional to $\sin2\theta_{13}$,
and so $\theta_{13}$ should not be too small otherwise the number of
events will be too small.

\begin{figure}
\centering
%\leavevmode
%\vspace{-2.cm}
\epsfxsize=4.0in
\epsffile{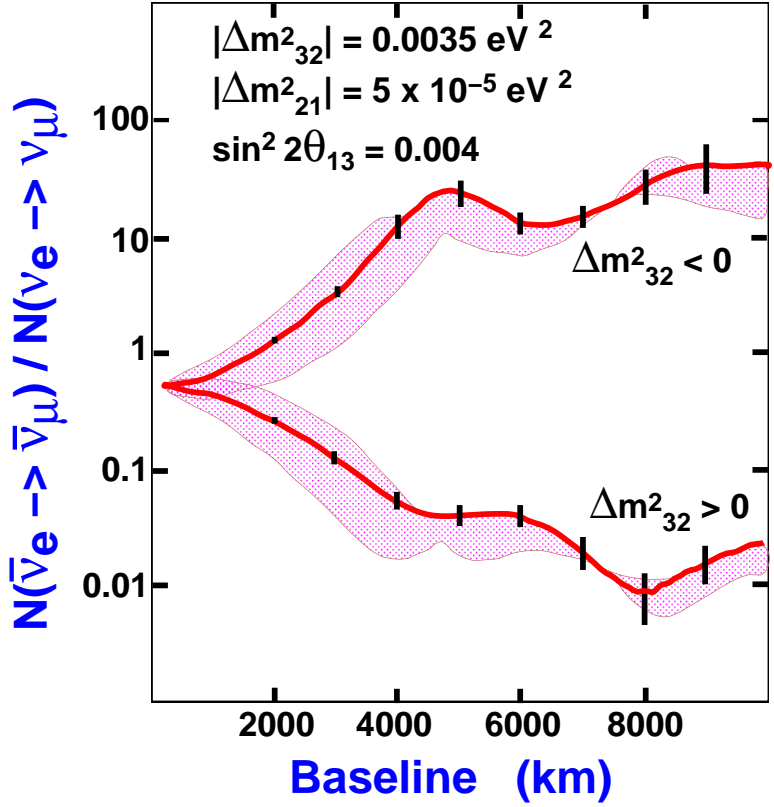}
%\vspace{2.cm}
\caption[]
{Measuring $\delta$ and sign($\Delta m_{32}^2$) at a Neutrino 
Factory. In this plot the bands correspond to varying $\delta$ between
0 and $\pi/2$, and  the error bars are estimated for a high
performance Neutrino Factory
(taken from ref.\cite{Geer}.)}
\label{fig5}
\end{figure}

Unfortunately life is not quite as simple as the above discussion
portrays. The Earth is made from matter and not anti-matter and so
CP will be violated by matter effects as the neutrino beam passes
through the Earth from the muon storage ring to the far detector.
For example the matter effects will modify the formulas for
$P(\nu_{e}\rightarrow \nu_{\mu})$ involving
$\theta_{13}$ and $\Delta m_{31}^2$ as follows:
\bea
\sin2\theta_{13} & \rightarrow &
\frac{\sin2\theta_{13}}
{\left( \frac{A}{\Delta
m_{31}^2}-\cos2\theta_{13}\right)^2+\sin^22\theta_{13}}
\nonumber \\
\Delta m_{31}^2 & \rightarrow &
\Delta m_{31}^2
\sqrt{\left( \frac{A}{\Delta
m_{31}^2}-\cos2\theta_{13}\right)^2+\sin^22\theta_{13}}
\label{matter}
\eea
where
\be
A=7.6\times 10^{-5}\rho E
\ee
where $\rho$ is the density of the Earth in gcm$^{-3}$ and $E$ is
the beam energy in GeV. The point is that
for $P(\bar{\nu}_{e}\rightarrow \bar{\nu}_{\mu})$
the sign of $A$ is reversed.
From one point of view
this is good news, since unlike the
vacuum oscillation formulae, $\Delta m_{31}^2$ enters linearly,
not quadratically, and so matter effects enable
the sign of the mass squared splitting to be determined in a
rather straightforward way, as shown in Figure \ref{fig5}.

However from the point of view of
measuring $\sin \delta$ it leads to complications since the
asymmetry in the rate in Eq.\ref{ACP} can get contributions from
both intrinsic CP violation and from matter induced CP violation,
and the measured asymmetry is a sum of the two effects
\be
A^{CP}=A^{\delta} + A^{matter}
\ee
Since both effects are by themselves rather small, it will be a
very difficult job to disentangle them, and the optimal
strategy continues to be studied \cite{NF1,NF2,NF3,CP}.
The optimal place to sit in order to observe CP violation
seems to be at the peak of $\sin (1.27\Delta m_{32}^2L/E)$ in
order to maximise $P^-$ according to Eq.\ref{P-}
(certainly we should avoid being at its node otherwise CP
violation vanishes). 
In order to do this efficiently
it may be desirable to have energy-tunable beams, and it is certainly
necessary to have a good understanding of the density profile
of the Earth. Assuming the LMA solution, the prospects for
measuring CP violation at a Neutrino Factory are good
as shown in Figure \ref{fig5}.

\section{More About the See-Saw Mechanism}

The basic idea of the see-saw mechanism \cite{seesaw} can most simply
be explained in the context of a single family of neutrinos,
where the Dirac mass coupling the left-handed neutrino
to the right-handed neutrino is $m_{LR}$ and the 
Majorana mass of the right-handed neutrino is $M_{RR}$.
The neutrino masses in the basis of left and right-handed neutrinos
can then be written as a mass matrix
\begin{equation}
\left(\begin{array}{cc}
0 & m_{LR} \\
m_{LR}^T & M_{RR} \\
\end{array}\right)
\end{equation}
where in this one family example $m_{LR}^T=m_{LR}$.
If $m_{LR}\ll M_{RR}$ then when this mass matrix is diagonalised
we find that one of the eigenstates may be identified
to good approximation with the left-handed neutrino and has
a very small Majorana mass $m_{LL}\approx m_{LR}M_{RR}^{-1}m_{LR}^T$.
For example if we take $m_{LR}=M_W$ and $M_{RR}=M_{GUT}$
then we find $m_{LL}\sim 10^{-3}$ eV which looks good for solar
neutrinos. Atmospheric neutrino masses would require 
a right-handed neutrino with a mass below the GUT scale.

Generalising the above result to three families of neutrinos
the light effective physical left-handed neutrino 
Majorana masses are now given by a three by three matrix
$m_{LL}$ which is given by
$m_{LL}=m_{LR}M_{RR}^{-1}m_{LR}^T$
where $m_{LR}$ is the Dirac neutrino mass matrix, typically
of the same magnitude as the charged lepton mass matrix, and
$M_{RR}$ is a heavy right-handed neutrino Majorana matrix
whose entries may be as large as the GUT scale. The eigenvalues
of $m_{LL}$ are the physical neutrino masses $m_1$, $m_2$, and $m_3$,
and, in the diagonal charged lepton mass basis, the matrix which
diagonalises $m_{LL}$ is the neutrino mixing matrix $U_{MNS}$.

Consider three left-handed neutrinos but only
two \footnote{This is purely for simplicity.
The following
argument also applies to three right-handed neutrinos but
is more cumbersome. Note that with three left-handed neutrinos
$m_{LL}$ is always a three by three matrix regardless
of the number of right-handed neutrinos.} right-handed neutrinos.
Then let us write the
Dirac mass matrix as
\begin{equation}
m_{LR}=
\left( \begin{array}{ccc}
0 & a & {d}\\
0 & b & {e}\\
0 & c & {f}
\end{array}
\right)
\end{equation}
where the notation LR means that the second and third columns
of $m_{LR}$ correspond to the second and third right-handed neutrinos.
The heavy Majorana mass matrix, assumed to be diagonal, is
\begin{equation}
M_{RR}=
\left( \begin{array}{ccc}
0 & 0 & 0    \\
0 & X & 0 \\
0 & 0 & Y
\end{array}
\right) 
\end{equation}
Then using the see-saw formula for the light effective Majorana
mass matrix $m_{LL}=m_{LR}M_{RR}^{-1}m_{LR}^T$, we find
the symmetric matrix,
\begin{equation}
m_{LL}
=
\left( \begin{array}{ccc}
{\frac{d^2}{Y}}+\frac{a^2}{X}
& {\frac{de}{Y}} +\frac{ab}{X}
& {\frac{df}{Y}}+\frac{ac}{X}    \\
.
& {\frac{e^2}{Y}} +\frac{b^2}{X}
& {\frac{ef}{Y}} +\frac{bc}{X}   \\
.
& .
& {\frac{f^2}{Y}} +\frac{c^2}{X}
\end{array}
\right)
\end{equation}

So far the discussion is fairly general, although we have
assumed a diagonal heavy Majorana mass matrix.
In order to account for the atmospheric
and solar neutrino data many models have been proposed 
based on the see-saw mechanism \cite{Models}.
One question which is common to all these models is
how to arrange for a large mixing angle involving the
second and third generation of neutrinos, without destroying the
hierarchy of mass splittings in Eq.\ref{R},
$R\equiv |\Delta m^2_{sol}|/|\Delta m^2_{atm}|$.
Assuming $\theta_{23}\sim \pi/4$ one might expect
two similar eigenvalues $m_2 \sim m_3$, and then the hierarchy
of scheme A in Figure \ref{fig3} seems rather unnatural.

One way to achieve a natural hierarchy is to 
suppose that the third right-handed neutrino contributions 
are much greater than the second right-handed neutrino contributions
in the 23 block of $m_{LL}$ \cite{SRHND},
\begin{equation}
\frac{(e^2,f^2,ef)}{Y}\gg
\frac{(b^2,c^2,bc)}{X}
\label{SRHND}
\end{equation}
This implies an approximately vanishing 23 subdeterminant,
\begin{equation}
det[m_{LL}]_{23}=
\left( \frac{e^2}{Y} +\frac{b^2}{X}\right)
\left( \frac{f^2}{Y} +\frac{c^2}{X}\right)
-\left( \frac{ef}{Y} +\frac{bc}{X}\right) ^2\approx 0
\label{det1}
\end{equation}
The 23 subdeterminant is also equal to the product
of the eigenvalues
\begin{equation}
det[m_{LL}]_{23}=m_{2}m_{3}
\label{det2}
\end{equation}
and hence from Eqs.\ref{det1},\ref{det2}
\begin{equation}
{m_{2}}/{m_{3}}\ll 1
\end{equation}
Thus the assumption in Eq.\ref{SRHND} that the third right-handed neutrino
gives the dominant contribution to the 23 block of
$m_{LL}$ naturally leads to a neutrino mass hierarchy.
This mechanism is called single right-handed neutrino dominance
(SRHND)\cite{SRHND}. 

In the limit that only a single right
handed neutrino contributes the determinant clearly exactly vanishes and we 
have $m_{2}=0$ exactly. However the sub-dominant contributions
from the second right-handed neutrino will give a small finite
mass $m_{2}\neq 0$ as required by the MSW solution to the
solar neutrino problem. 

Developing the simple example above a little further,
assuming SRHND as discussed above,
we may obtain simple analytic estimates for the
neutrino masses \cite{SRHND2}:
\begin{equation}
m_{1}=0,
\label{m1}
\end{equation}
\begin{equation}
m_{2}\sim \frac{(b-c)^2}{X}
\label{m2}
\end{equation}
\begin{equation}
m_{3}\approx \frac{(d^2+e^2+f^2)}{Y}
\label{m3}
\end{equation}
Note that $m_3$ ($m_2$) is determined by 
parameters associated with the dominant (subdominant)
right-handed neutrino.
Given the SRHND assumption in Eq.\ref{SRHND}
we see that we have a generated a hierarchical spectrum
$m_1\ll m_2\ll m_3$ as in scheme A of Figure \ref{fig3}.

The mixing angles may also be estimated to be \cite{SRHND2}
\begin{equation}
\tan \theta_{23} \sim {\frac{e}{f}}, 
\end{equation}
\begin{equation}
\tan \theta_{13} \sim {\frac{d}{\sqrt{e^2+f^2}}}
\end{equation}
\begin{equation}
\tan \theta_{12} \sim { \frac{\sqrt{2}a}{b-c}}
\end{equation}

For example by a suitable choice of parameters $e=f\gg d$ 
and $a\sim (b-c)$ it is possible to
have large $\theta_{12}$ suitable for the LMA or LOW solution and a maximal
$\theta_{23}$ suitable for atmospheric oscillations, while maintaining
a small $\theta_{13}$ consistent with the CHOOZ constraint.
Note that the 
hierarchical masses in Eqs.\ref{m1},\ref{m2},\ref{m3}  
are controlled by the SRHND condition Eq.\ref{SRHND}
and the hierarchy is unaffected by the large angle conditions above. 
This is due to the approximately vanishing 23 subdeterminant
of $m_{LL}$, and the underlying physical mechanism responsible 
is SRHND. Furthermore the low energy predictions are
quite stable under radiative corrections, and a recent renormalisation
group analysis shows corrections of only a few per cent \cite{SRHNDRGE}.

These ideas may be implemented 
in flavour models based on a $U(1)$ family
symmetry but the predictions for different models are uncertain
due to unknown coefficients multiplying expansion
parameters in the neutrino matrices \cite{SRHND}.
Within such a framework
the best one can do is scan over the unknown coefficients,
and plot distributions in the predicted quantities \cite{HK}.
See-saw models with SRHND are clearly seen to give distributions
in $R$ (defined in Eq.\ref{R})
which peak at smaller values than other models,
so the idea of SRHND is testable by a future measurement
of $R$. The result of such scans shows that
even without the SRHND mechanism it is possible
to achieve mild hierarchies corresponding to say $R\sim 0.1$
in a perfectly natural way. However if experiment
tells us that $R\leq 10^{-2}$, which may be the case
for LMA and is certainly the case for LOW, 
then an accidental hierarchy of this
magnitude looks increasingly
unlikely, and in this case the simplest interpretation
might be SRHND.

\section{Conclusion}
Despite the impressive progress we still in reality know
very little about the neutrino spectrum. For example,
the pattern of masses could be hierarchical, inverted or degenerate.
According to the recent Super-Kamiokande results 
which favour the LMA solution with active neutrinos,
the most likely scenario is approximate bimaximal mixing
$\theta_{23}\approx \pi/4$, $\theta_{12}\approx \pi/4$, 
$\theta_{13}\approx 0$ \cite{bimaximal},
\begin{equation}
U_{MNS}\approx
\left( \begin{array}{ccc}
\frac{1}{\sqrt{2}} & \frac{1}{\sqrt{2}} & 0\\
-\frac{1}{2} & \frac{1}{2} & \frac{1}{\sqrt{2}}\\
\frac{1}{2} & -\frac{1}{2} & \frac{1}{\sqrt{2}}
\end{array}
\right)
\end{equation}
However this interpretation is far from certain at this stage.
The solar angle might still be very small, as in the SMA solution.
There could still be a fourth (sterile) neutrino state.
One or more of the solar experiments could be wrong;
for example if the Homestake Chlorine data is removed then
the allowed parameter space opens up significantly,
and the distinction between the LMA, LOW and VAC solutions
becomes less pronounced. 
Then even trimaximal mixing with a large CHOOZ angle cannot strictly
be excluded \cite{trimaximal}.
Finally perhaps we are not
seeing oscillations, but some energy independent
suppression for example due to the
effects of large extra dimensions, or perhaps some other
non-standard effect. 
Fortunately the answers to these
questions will be found by the forthcoming generation
of neutrino experiments. As Super-Kamiokande leaves centre stage,
other experiments such as those
discussed in section 5 are set to enter the spotlight.
The new generation of solar neutrino experiments
will be able to confirm solar neutrino oscillations,
and the LBL experiments will be able to confirm atmospheric
oscillations. Together they will either confound the sceptics,
or else dumbfound the believers, and it is a brave person
who would predict that there will be no more surprises in store.

For example at the time of writing
we are awaiting the first results from the first year of operation
of SNO with great anticipation.
The first year of data will consist of a measurement of the
charged current scattering rate from deuterium, 
which will provide a clean measurement of the electron neutrino flux.
By contrast Super-Kamiokande measures the elastic scattering
rate from electrons, which includes charged and neutral
current contributions. By comparing the charged current suppression
measured by SNO to the elastic scattering suppression measured
by Super-Kamiokande, it will be possible to infer the neutral
current component of the elastic scattering
observed by Super-Kamiokande, and hence test the hypothesis
that electron neutrinos oscillate
into muon and tau neutrinos (rather than sterile neutrinos), even
before SNO measures the neutral current rate directly \cite{SNO2}.

Suppose that neutrino masses are confirmed - so what?
I have heard it suggested that neutrino masses are just
a trivial extension to the Standard Model and hardly worth mentioning,
and that learning the remaining MNS parameters is nothing
more than stamp-collecting. It is true that one can simply 
add three light right-handed neutrinos (without Majorana masses) and
couple them to the three left-handed neutrinos via Standard Model-like 
Yukawa couplings, to describe the neutrino spectrum. 
However a third family 
neutrino mass of 0.05 eV is much smaller than the tau lepton
mass of 1777 MeV, and would require a tau neutrino Yukawa coupling 35 billion
times smaller than the tau lepton Yukawa coupling, 
which looks rather unnatural when compared to the
top-bottom quark mass ratio of about 40.
On the other hand
light right-handed neutrinos are not protected from receiving
Majorana masses by any gauge symmetry, so it is perfectly
natural for them to be much heavier than the weak scale,
in which case naturally small neutrino masses can arise
via the see-saw mechanism. In this case the scale of the heavy
right-handed neutrino mass will be associated with new physics
such as GUTs or string theory. 

The origin of the heavy
right-handed neutrino Majorana mass scale is in itself
very interesting, and it may be useful to classify
theories in terms of whether the right-handed neutrinos
carry gauge quantum numbers at high energy
(as in gauged $SU(2)_R$ models or $SO(10)$)
or are gauge singlets even at high energy (as in $SU(5)$).
For example if the right-handed neutrinos tranform
as part of a gauged $SU(2)_R$ doublet, then anomaly cancellation
predicts three right-handed neutrinos, and neutrino masses are inevitable.
However if the ``right-handed neutrinos'' are really gauge
singlets at high energies then the number and mass of such states
is undetermined. Additional gauge singlets are also quite
plausible in the gauged $SU(2)_R$ case. In general
there may be several gauge singlets, with a sequence of
vacuum expectation values and scales, making the see-saw mechansim
quite a complicated problem, and this is typical in string 
theories. In all these cases some simplification might arise if
one of the singlets plays the dominant role in the see-saw
mechanism as in SRHND. A further window
into high energy right-handed neutrino physics is provided
by Leptogenesis \cite{lepto}.
The heavy right-handed
neutrinos may be produced in the early universe, and decay out
of equilibrium into higgs plus leptons, producing a lepton asymmetry
as a result of CP violation in the lepton sector, which is then
reprocessed by sphaleron interactions into the observed baryon number
of the universe \cite{lepto}.
\footnote{Note that the amount of CP violation in the lepton sector
necessary to achieve leptogenesis does not imply that the
Dirac phase $\delta$ will be large enough to be 
measurable at a Neutrino Factory, but it may be \cite{HK2}.}
The alternatives to the see-saw mechanism, large extra dimensions
or R-parity violating supersymmetry, are hardly less exciting.

It is important to determine
the detailed pattern of neutrino masses and mixing angles,
since some day this will be related to the quark masses and mixing
angles in the framework of an all-encompassing theory
\cite{Models,KO}.
To determine $\theta_{13}$, the pattern of neutrino masses
and the CP violating Dirac phase may require a Neutrino Factory.
Without knowing the detailed spectrum, we shall most likely
never know if we have the correct theory since we shall not know the
low energy observables that we seek to explain.
This is of course the normal way that science progresses,
first the phenomena then the theory which lies underneath the
observations. Eventually there is no doubt that knowing the 
neutrino masses and mixing angles will help to unlock the puzzle
of all fermion masses and mixing angles, but we may have to wait
for further direct information about new physics, as
for example would be the case if superpartners are found at colliders,
before all the pieces of the puzzle can be assembled.
For example there is an interesting interplay between the
slepton mass matrix and the see-saw parameters, in the
framework of supersymmetry \cite{sacha}.

Neutrino oscillation physics is arguably the fastest developing
area of particle physics at the moment.
For example since 1998 more papers have appeared containing the word
neutrino(s) than the word quark(s) in the title.
With progress in this
area showing no sign of slowing up as the new experiments
successively come on line, this is a trend which seems likely to continue.

\begin{center}
{\bf Acknowledgements}
\end{center}
It is a pleasure to thank the organisers of the IPPP Workshops
on Physics at a Future Neutrino Factory.
I would also like to thank my colleagues
on the U.K. Neutrino Factory Panel for their help
in preparing this document.
In particular I would like to thank David Wark for his
help in preparing section 5 of this review and 
Ken Long for reading and commenting on the manuscript.
I would also like  to thank Martin Hirsch and Alan Martin for reading the
manuscript and for useful suggestions.
I also thank Martin Hirsch for numerous insightful
discussions about neutrino physics.
Finally I would like to thank PPARC for the support of a Senior Fellowship.

\section{Addendum: SNO News is Good News}
Since this review was completed SNO have announced their first
results \cite{SNO3}. The charged current (CC) reaction on deuterium
is sensitive exclusively to $\nu_e's$,  while the elastic scattering
(ES) off electrons also has a small sensitivity to $\nu_{\mu}'s$ and
$\nu_{\tau}'s$. SNO measures the flux of $\nu_e's$ from 
$^8B$ decay measured by the CC reaction to be:
\begin{equation}
\phi^{CC}_{SNO}(\nu_e)=1.75\pm 0.07^{+0.12}_{-0.11}\pm 0.05
\times 10^6 {\rm cm^{-2}s^{-1}}
\end{equation}
SNO also measures the flux of all neutrinos from 
$^8B$ decay by the ES reaction to be:
\begin{equation}
\phi^{ES}_{SNO}(\nu_x)=2.39\pm 0.34^{+0.16}_{-0.14}
\times 10^6 {\rm cm^{-2}s^{-1}}
\end{equation}
which is in agreement with the precision ES flux measured 
by Super-Kamiokande
\begin{equation}
\phi^{ES}_{SK}(\nu_x)=2.32\pm 0.03^{+0.08}_{-0.07}
\times 10^6 {\rm cm^{-2}s^{-1}}
\end{equation}
The difference between the precision SK ES flux and the
SNO CC flux measurement is
\begin{equation}
\phi^{ES}_{SK}(\nu_x) - \phi^{CC}_{SNO}(\nu_e)
=0.57\pm 0.17 \times 10^6 {\rm cm^{-2}s^{-1}}
\end{equation}
or 3.3 $\sigma$.
The ratio of the SNO CC $^8B$ flux to the latest standard solar
model prediction is 
\begin{equation}
\frac{CC(^8B)_{SNO}}{BP2001}=0.347\pm 0.029
\end{equation}
to be compared to the ratio of the SK ES
$^8B$ flux to the latest standard solar
model prediction  
\begin{equation}
\frac{ES(^8B)_{SK}}{BP2001}=0.451^{+0.017}_{-0.015}
\end{equation}

Clearly the CC ratio is significantly smaller than the 
ES ratio. This immediately disfavours oscillations of $\nu_e's$
to sterile neutrinos which would lead to a diminished flux
of electron neutrinos, but equal CC and ES ratios.
On the other hand the different ratios are consistent
with oscillations of $\nu_e's$ to active neutrinos $\nu_{\mu}'s$ and
$\nu_{\tau}'s$ since this would lead to a larger ES rate since this has
a neutral current component.
If we identify the heavy atmospheric neutrino mass as being approximately
$\frac{\nu_{\mu}+\nu_{\tau}}{\sqrt{2}}$ then the electron neutrino may
oscillate into the lighter orthogonal combination,
\be
\nu_e \rightarrow \frac{\nu_{\mu}-\nu_{\tau}}{\sqrt{2}}
\ee
By comparing the SK ES rate to the SNO CC rate, SNO infers a flux
of $\nu_{\mu}'s$ and $\nu_{\tau}'s$
\begin{equation}
\phi^{SNO}_{SK}(\frac{\nu_{\mu}-\nu_{\tau}}{\sqrt{2}})
=3.69\pm 1.13
\times 10^6 {\rm cm^{-2}s^{-1}}
\end{equation}
The total flux of active $^8B$ neutrinos is then determined to be
\begin{equation}
\phi^{CC}_{SNO}(\nu_e)+
\phi^{SNO}_{SK}(\frac{\nu_{\mu}-\nu_{\tau}}{\sqrt{2}})
=5.44\pm 0.99
\times 10^6 {\rm cm^{-2}s^{-1}}
\end{equation}
which is in excellent agreement with the Standard Solar Model
reference $^8B$ flux of $5.05 \pm 1.0\times 10^6 {\rm cm^{-2}s^{-1}}$
\cite{BPB}. 
The SNO analysis
is nicely consistent with both the hypothesis that electron neutrinos from
the Sun oscillate into other active flavours, and with the
Standard Solar Model prediction.
In other words there is no longer any solar neutrino problem: we
have instead solar neutrino mass!

One can go further and ask which particular type of active neutrino
oscillation best fits the SNO data?
The results of recent global
analyses, performed just days after the SNO result \cite{recent}
show that the most favoured solutions are either the LMA
or the LOW  solutions, with a slight preference for the
LMA solution. The SMA solution is disfavoured at about
3 $\sigma$, while the VAC regions are disfavoured at various
confidence levels. It now looks like a two horse race between
LMA and LOW, with LOW very much the dark horse,
having gained a lot of ground over the past few years.

Out of the two preferred solutions, LMA or LOW, the LMA solution
will be tested by KamLAND in about a year from now.
In the meantime SNO has now added the salt which will enable
a direct measurement of the neutral current flux, and results which
will confirm its first findings are expected on a similar timescale.

What of the theoretical implications of these results?
The large extra dimension approaches to small neutrino masses
with right-handed neutrinos in the bulk tend to resemble
neutrino oscillations into sterile neutrinos and the SMA solution
\cite{LED}. The R-parity violating approaches based on
minimal supergravity assumptions commonly
yield SMA solutions \cite{RPV}. On the other hand the see-saw
mechanism \cite{seesaw} combined with SRHND \cite{SRHND2}
can readily yield LMA or LOW solutions, as the discussion
in section 7 shows. Indeed the LOW solution involving
an extreme hierarchy of neutrino masses $R\ll 1$, makes
SRHND almost mandatory. So, the see-saw mechanism is looking very
good in the light of the recent SNO result, and this implies a large
mass scale in  physics associated with the right-handed neutrinos,
and the possibility of leptogenesis. The large mass scale may be
related to string theory or GUTs, and the presence of such a large
mass scale may be taken to imply TeV scale supersymmetry in order
to stabilise the weak scale. Supersymmetry in turn may provide a
window into the see-saw parameters via the slepton mass
matrix and lepton flavour violation \cite{sacha}.
If this scenario turns out to be the way things are, then
the accurate experimental measurement of the neutrino parameters will 
provide important information which will be useful
for the determination of the correct
high energy theory.

\end{document}